\documentclass[12pt]{article}
\usepackage{amsmath}
\usepackage{amssymb}

\newcommand{\be}{\begin{equation}}
\newcommand{\bea}{\begin{eqnarray}}
\newcommand{\ee}{\end{equation}}
\newcommand{\eea}{\end{eqnarray}}
\def\theequation{\arabic{section}.\arabic{equation}}
\newcommand{\hp}{\hat{p}}

\newcommand{\bn}[2]{\left(\begin{array}{c} #1\\ #2
\end{array}\right)}

\begin{document}
\topmargin -1cm \oddsidemargin=0.25cm\evensidemargin=0.25cm
\setcounter{page}0
\renewcommand{\thefootnote}{\fnsymbol{footnote}}
\begin{titlepage}
\begin{flushright}
LTH-923
\end{flushright}
\vskip .7in
\begin{center}
{\Large \bf On the Structure of  Quartic Vertices for Massless Higher
Spin Fields on Minkowski Background } \vskip .7in {\large Paul
Dempster$^a$\footnote{e-mail: {\tt paul.dempster@liv.ac.uk }} and Mirian
Tsulaia$^{b}$\footnote{e-mail: {\tt mirian.tsulaia@gmail.com} }} \vskip .2in {$^a$ \it
Department of Mathematical Sciences, University of Liverpool,
Liverpool, L69 7ZL, United Kingdom} \\
\vskip .2in { $^b$ \it Centre of Particle Physics, Institute for Theoretical Physics, Ilia State University, 0162, Tbilisi,
Georgia }\\

\begin{abstract}

We consider in detail the structure of quartic vertices for massless higher spin fields
on Minkowski background, and  study the consistency conditions imposed
on cubic and quartic interactions by symmetries of the S--matrix.
We discuss the possibility of generalizing the construction of  quartic vertices to
 ${\cal D}$-dimensional anti-de Sitter space.

\end{abstract}

\end{center}

\vfill

\end{titlepage}

\tableofcontents

\section{Introduction}
One of the most important problems in higher spin gauge theory
(see \cite{Vasiliev:1999ba}--\cite{Tsulaia:2012rb} for reviews of different aspects of the subject)
on Minkowski background is a problem of consistency
of cubic and higher order interactions. It is known that the
 interaction between massless higher spin fields on a flat background, unlike the one on $AdS_{\cal D}$ space which naturally bypasses the
Coleman--Mandula no--go theorem, is considered to be more problematic. However,
the problem of constructing  cubic interaction vertices
for massless higher spin fields  on  flat space--time background has been extensively studied in
\cite{Bengtsson:1987jt}--\cite{Manvelyan:2010je}  and nontrivial solutions for cubic vertices have been found.
An important question, however, is if one can build a consistent perturbation theory
for massless higher spin fields on Minkowski space-time in order to obtain a nontrivial
S--matrix, since
 the most formidable problems in higher spin gauge theory on flat background start from four point interactions.

The problem of consistency
of  four point functions for massless higher spin fields on Minkowski background
was addressed in \cite{Fotopoulos:2010ay} (see also \cite{Sagnotti:2010at},
\cite{Polyakov:2010sk}, \cite{Taronna:2011kt} for the construction and extensive discussion of quartic interactions as well as \cite{Bengtsson:2006pw}
for a general strategy). The direct evaluation of four point functions in terms of Feynman diagrams using the results
of \cite{Bekaert:2009ud}
as well as analysis based on BCFW recursion relations suggested that consistency of the interacting theory of massless higher spin
fields on a flat background  might require an addition of extra nonlocal or composite objects, such as stringy Pomerons
\cite{Brower:2006ea}.

The aim of the present paper is to perform a detailed study of the structure of quartic vertices using the comprehensive results
for cubic vertices obtained in \cite{Bengtsson:1987jt}--\cite{Sagnotti:2010at} and the further analysis of consistency conditions for corresponding four point functions, in the spirit of \cite{Fotopoulos:2010ay}.
In Section \ref{QUAR}    we collect the main equations which are needed for the construction
of cubic and quartic interaction vertices for massless higher spin fields on flat background.
Further we
we describe an explicit form of cubic vertices considered earlier in \cite{Sagnotti:2010at} and
of their off-shell BRST invariant extension given in \cite{Fotopoulos:2010ay}.
The vertices given in \cite{Fotopoulos:2010ay} are products of three exponentials, each of them being separately BRST invariant,
because their arguments $\Delta_i$, $i=1,2,3$ are BRST invariant expressions  themselves.
For the sake of simplicity we prefer to work with on--shell vertices
i.e. with the on--shell form of $\Delta_i$, when the BRST invariance
is maintained   using of the free equations of motion,
rather  with identically BRST invariant vertices.
Since any function of $\Delta_i$ is BRST invariant as well and therefore is  a priori a valid
cubic vertex,
in  Section \ref{QV} we allow an arbitrary dependence of the cubic vertex on $\Delta_i$
and find an explicit form of the corresponding quartic vertices.
However,  the existence of explicit solutions for quartic vertices still does not guarantee
that the corresponding four point amplitudes are nontrivial or consistent.
In order to check if one can have a consistent interacting theory of massless higher spin particles
on four dimensional Minkowski background in Section \ref{SM}
 we use a ``four particle test" for consistency of four point amplitudes introduced in \cite{Benincasa:2007xk}, which in turn is based
on BCFW recursion relations \cite{Britto:2004ap}--\cite{Britto:2005fq}.
We apply this test for the simplest situation of a four point tree level amplitude, where external particles
are real scalars, whereas  intermediate propagators correspond to an infinite number of massless higher spin fields.
We find that, similarly to what was obtained in \cite{Fotopoulos:2010ay}, although the
explicit form of quartic vertices can be found,
the four point functions do  not pass the ``four particle test'' of \cite{Benincasa:2007xk}.
  Further,
in  Section \ref{QADS} we discuss a general strategy for computations of quartic vertices in ${\cal D}$--dimensional $AdS_{\cal D}$ spaces and  present identities which are needed for these computations.
 We conclude with a summary of our results and some comments.
Some technical details of the calculations are collected in the Appendix.

\setcounter{equation}0 \section{Basic equations: quartic Lagrangians}\label{QUAR}

Let us briefly recall the BRST approach for constructing quartic Lagrangians \cite{Fotopoulos:2010ay}.
A detailed review of the formalism for cubic Lagrangians can be found in \cite{Fotopoulos:2008ka}.

In order to describe cubic Lagrangians for higher spin fields one takes three copies of
Hilbert spaces spanned by oscillators
\begin{equation}
[\alpha_\mu^i, \alpha^{j+}_\nu]= \eta_{\mu \nu}\delta^{ij}, \quad  \{c^i_0, b^j_0 \} = \{c^i, b^{j+} \}= \{c^{i+}, b^j \}=\delta^{ij}.
\end{equation}
The corresponding cubic Lagrangian has a form
\begin{equation} \label{LIBRSTQ-3}
{L}  =  \sum_{i=1}^3 \int d c_0^i \langle \Phi_i |\, Q_i \,|\Phi_i \rangle \ \\ \nonumber
  + g (\int dc_0^1 dc_0^2  dc_0^3 \langle \Phi_1|
\langle \Phi_2|\langle \Phi_3||V_3 \rangle + h.c),
\end{equation}
where $g$ plays the role of a coupling constant.
For massless bosonic higher spin fields
one considers the nilpotent BRST charge
\begin{equation} \label{brst}
Q^i =
c_0^i{l}_0^i +c^i l^{i+}+c^{i+} l^i  - c^{i+}c^ib^i_0, \quad i=1,2,3,4,
\end{equation}
where we have introduced the d'Alembertian, the divergence and the symmetrized exterior derivative operators\footnote{We use notation $A \cdot B = A^\mu B_\mu$.}
\begin{equation} \label{lapl} l_0^i \ = \
  p^i \cdot p^i\,, \quad l^i =  \alpha^{ i} \cdot p^i, \quad
l^{i +} =  \alpha^{i +} \cdot p^i\, ,
\end{equation}
and $p_\mu = -i \partial_\mu$.
The Lagrangian (\ref{LIBRSTQ-3}) is invariant up to the first order in $g$
 under  nonlinear transformations with  parameters $|\Lambda_i \rangle$
\begin{equation}
\delta | \Phi_i \rangle = Q_i |\Lambda_i \rangle
- g
\int
dc_0^{i+1} dc_0^{i+2}[(  \langle \Phi_{i+1}|\langle \Lambda_{i+2}|
+\langle \Phi_{i+2}|\langle \Lambda_{i+1}|) |V_3 \rangle],
\end{equation}
provided the cubic vertex $| V_3 \rangle$ is  invariant under the nilpotent BRST charge $Q=Q_1+Q_2+Q_3$.

As the next step one can generalize this construction for a system which contains quartic interactions.
The corresponding Lagrangian for interacting higher spin fields up to the quartic order in the fields has a general form
\begin{eqnarray} \label{LIBRSTQ}
{L} & = & \sum_{i=1}^4 \int d c_0^i \langle \Phi_i |\, Q_i \,|\Phi_i \rangle \ \\ \nonumber
  &&+ g( \int dc_0^1 dc_0^2  dc_0^3 \langle \Phi_1|
\langle \Phi_2|\langle \Phi_3||V_3 \rangle
+\int dc_0^1 dc_0^2  dc_0^4 \langle \Phi_1|
\langle \Phi_2|\langle \Phi_4||V_3 \rangle \\ \nonumber
&&+
\int dc_0^2 dc_0^3  dc_0^4 \langle \Phi_2|
\langle \Phi_3|\langle \Phi_4|V_3 \rangle
+
\int dc_0^1 dc_0^3  dc_0^4 \langle \Phi_1|
\langle \Phi_3|\langle \Phi_4||V_3 \rangle
+ h.c) \\ \nonumber
&& + g^2 (\int dc_0^1 dc_0^2  dc_0^3  dc_0^4  \langle \Phi_1| \langle \Phi_2|\langle \Phi_3| \langle \Phi_4|  |V_4 \rangle + h.c).
\end{eqnarray}
Let us require  the Lagrangian (\ref{LIBRSTQ})
to be invariant under the following nonlinear gauge transformations
\begin{equation}
\delta | \Phi_i \rangle= (\delta_0 + \delta_1 + \delta_2) | \Phi_i \rangle,
\end{equation}
where
\begin{equation}\label{BRSTIGT1Q0}
\delta_0 | \Phi_i \rangle  =  Q_i | \Lambda_i \rangle,
\end{equation}
\begin{eqnarray} \label{BRSTIGT1Q1}
\delta_1 | \Phi_i \rangle &=&
- g(
\int
dc_0^{i+1} dc_0^{i+2}[(  \langle \Phi_{i+1}|\langle \Lambda_{i+2}|
+\langle \Phi_{i+2}|\langle \Lambda_{i+1}|) |V_3 \rangle]+ \\ \nonumber
&& \int
dc_0^{i+2} dc_0^{i+3}[(  \langle \Phi_{i+2}|\langle \Lambda_{i+3}|
+\langle \Phi_{i+3}|\langle \Lambda_{i+2}|) |V_3 \rangle] + \\ \nonumber
&&
\int
dc_0^{i+1} dc_0^{i+3}[(  \langle \Phi_{i+1}|\langle \Lambda_{i+3}|
+\langle \Phi_{i+3}|\langle \Lambda_{i+1}|) |V_3 \rangle]), \\ \nonumber
\end{eqnarray}
\begin{eqnarray}  \label{BRSTIGT1Q2-1} \nonumber
\delta_2 | \Phi_i \rangle &=&
{(-1)}^ig^2\int dc_0^{i+1}
dc_0^{i+2} dc_0^{i+3}[(  \langle \Phi_{i+1}|   \langle \Phi_{i+2}|\langle \Lambda_{i+3}| +
 \langle \Phi_{i+1}|   \langle \Phi_{i+3}|\langle \Lambda_{i+2}|+ \\
&& \langle \Phi_{i+2}|   \langle \Phi_{i+3}|\langle \Lambda_{i+1}|)
 |V_4 \rangle].
\end{eqnarray}
The Lagrangian (\ref {LIBRSTQ}) is invariant
up to zeroth order in the coupling constant $g$, i.e. under
 (\ref{BRSTIGT1Q0}),
  since each separate BRST charge $Q_i$ is nilpotent.
  Further, the Lagrangian is
 invariant up to first order in the coupling constant, g,  i.e. under
\eqref{BRSTIGT1Q0} and (\ref{BRSTIGT1Q1}), provided
\be \label{brstinv}
(Q_i+Q_j+Q_k)|V_3 \rangle=0, \quad i \neq j  \neq k.
\ee
At the same time the equation (\ref{brstinv}) ensures the closure of the algebra of gauge transformations at first order
in the coupling constant $g$.

A quartic vertex $| V_4 \rangle$ is determined  from the requirement of cancelation
of the terms of order $g^2$ in the variation of the Lagrangian (\ref{LIBRSTQ}).
After introducing the notation
\begin{equation}
\int d c_0^i\;{}_{a, b, i}\langle V_3 || V_3 \rangle_{\alpha, \beta, i} = {\cal V}(a, b ;\alpha, \beta),
\end{equation}
and the fact that
\begin{equation}\label{223}
[\int dc_0^i  d c_0^j \;{}_i\langle A |\;{}_j\langle B||V_3 \rangle_{i,j,k}]^\dagger= \int dc_0^i  d c_0^j \;{}_{i,j,k}\langle V_3||B \rangle_i | A\rangle_j,
\end{equation}
one finally gets the equation for determining the quartic vertex
\begin{eqnarray} \label{quartbrst} \nonumber
\frac{1}{3}(Q_1+Q_2+Q_3+Q_4)|V_4 \rangle_{a,b, \alpha, \beta} &=& ({\cal V}(a, b ;\alpha, \beta) + {\cal V}(b, \alpha ; a, \beta) - (a \leftrightarrow b))
\\
&&+ ({\cal V}(\alpha, a ;b, \beta) - (a \leftrightarrow \alpha)).
\end{eqnarray}
Again, the ``generalized Jacobi identities" (\ref{quartbrst}) ensure the closure of the algebra of gauge transformations at order
$g^2$ in the coupling constant.

One can further simplify the discussion for quartic vertices by considering on--shell
 vertices instead of off--shell ones. First let us note that,
 because of the presence of ghost variables, the Lagrangian contains auxiliary degrees of freedom.
 However, one can fix the gauge in order to eliminate all nonphysical degrees of freedom (see for example \cite{Fotopoulos:2008ka}), i.e.
obtain
 $|\Phi_i \rangle = |\phi_i \rangle $, where the field $|\phi_i \rangle $ depends only on the oscillators $\alpha^{i+}$.
 As a result the free part of the Lagrangian describes the propagation of fields
  with the spins\footnote{Let us note that an irreducible higher spin mode can be described by adding a
zero trace condition to the higher spin field,
An entire discussion of the present Section can be reformulated for the irreducible massless or massive higher spin modes
without any changes. The only difference is that one has to use the more complicated form of the BRST charge
 \cite{Pashnev:1998ti}--\cite{Buchbinder:2005ua} (see also \cite{Francia:2002pt} for  nonlocal formulation of dynamics of massless higher spin fields
 and \cite{Buchbinder:2007ak} for unconstrained ``quartet" formulation of massless and massive higher spin fields)
  which gives the free equations describing irreducible  higher spin modes after solving its cohomologies.}
$s, s-2,s-4,...,1/0$. These fields obey the equation of motion
  \begin{equation}\label{EOM}
  l^i_0|\phi_i \rangle=l^i|\phi_i \rangle=0.
  \end{equation}
  The equations \eqref{EOM} are still  invariant under gauge transformations with the parameter
  $|\lambda_i \rangle$
  \begin{equation}\label{gauget}
  \delta |\phi_i \rangle = l^{i+} |\lambda_i \rangle, \quad l^i_0|\lambda_i \rangle=l^i|\lambda_i \rangle=0,
  \end{equation}
  where no summation over $i$ in the equations (\ref{EOM})--(\ref{gauget}) is assumed.
 As a result cubic and quartic interactions have the form
 \be \label{intred}
\langle \phi_1|
\langle \phi_2|\langle \phi_3||V_3 (\alpha^+) \rangle, \quad \textrm{and} \quad \langle \phi_1|
\langle \phi_2|\langle \phi_3| \langle \phi_4 ||V_4 (\alpha^+) \rangle,
\ee
where $|V_3 (\alpha^+) \rangle$  and  $|V_4 (\alpha^+) \rangle$ are the parts of the cubic and quartic vertices which depend only on
the $\alpha^{i+}$ oscillators.
On the other hand,
  an on--shell version of equation (\ref{quartbrst})
 gives
\begin{eqnarray} \label{quartons}
&&\sum_{i=1}^4 l^{i+4} \langle \phi_{i+1}| \langle \phi_{i+2}|\langle \phi_{i+3}| \langle \lambda_{i+4}| |V_4 \rangle = \\ \nonumber
&&g^2[ \langle V_3(1,2,3)| |\phi_2 \rangle |\phi_3 \rangle]
[\langle \lambda_{2'}|
\langle \phi_{3'}||V_3^\prime (1,2',3')(\alpha^+) \rangle]
+  \textrm{permutations},  \nonumber
\end{eqnarray}
where $|V_3^\prime (1,2',3')(\alpha^+) \rangle$ denotes a part of the first derivative of
the cubic vertex with respect to the combination $c^{i+}b_0^j$ which depends only on the oscillators
$\alpha^{i+}$. Similarly
the BRST invariance condition (\ref{brstinv}) of the cubic vertex
implies that
 \be \label{brstinvons}
(l^i+l^j+l^k)|V_3 \rangle=0, \quad i \neq j  \neq k.
\ee

   The equations for on--shell vertices are simpler to analyze because of their more compact
form compared to their off--shell completions.
For this reason we shall mainly consider on--shell vertices.
Let us first consider an off--sell cubic vertex given in   \cite{Fotopoulos:2010ay}.
 This vertex can be written as a product of
three vertices
\begin{equation} \label{TOTAL}
V_3 |0\rangle_{123} =
e^{\Delta_1+\Delta_2+\Delta_3 }|0\rangle_{123},
\end{equation}
where each of the expressions $\Delta_i$ is separately BRST invariant.
According to the discussion above, in order to obtain an on--shell formulation \cite{Sagnotti:2010at} one needs to keep
 only  the parts of
$\Delta_i$ which do not depend on ghosts and the parts of $\Delta_i$ which are linear in the combination
$c^{i+}b_0^j$. Using the explicit form of $\Delta_i$ given in \cite{Fotopoulos:2010ay}
one obtains\footnote{The easiest way to obtain these expressions  from the vertex given in \cite{Fotopoulos:2010ay} is as follows.
First note that one can  set the diagonal components of the matrices
$Y_{ij}$ and $Z_{ij}$ and $X^{(m)}_{rstu}$equal to zero using  momentum conservation
and removing a BRST-trivial part.
In this way one obtains a ``truncated" solution.
For $\Delta_1$ one has
$Y_{12}=Y_{23}=Y_{31}=a_1, \quad Y_{21}=Y_{32}=Y_{13}=-a_1,$ and
$Z_{12}=Z_{23}=Z_{31}=-a_1, \quad Z_{21}=Z_{32}=Z_{13}=a_1$
whereas for $\Delta_3$ one has
${\tilde X}^{(1)}_{1231}=a_3, \quad {\tilde X}^{(1)}_{1232}=-a_3, \quad
{\tilde X}^{(2)}_{1231}=-2a_3, \quad
{\tilde X}^{(2)}_{2132}=2a_3,$ and
${\tilde X}^{(3)}_{1231}=-a_3, \quad {\tilde X}^{(3)}_{1232}=a_3, \quad
{\tilde X}^{(4)}_{1231}=a_3, \quad
{\tilde X}^{(4)}_{1232}=-a_3, \quad {\tilde X}^{(4)}_{1233}=a_3$. After that one takes  $a_1=a_3$.
 }
\begin{equation} \label{deltas-1}
\Delta_1= Y^+_{\alpha}+ Y^+_{gh.},  \quad \Delta_3= X^+_{\alpha}+ X^+_{gh.},
\end{equation}
\begin{equation} \label{deltas-2}
\Delta_2 =  \frac{a_2}{2}(\alpha^{1+} \cdot \alpha^{1+}+\alpha^{2+} \cdot \alpha^{2+}+\alpha^{3+} \cdot \alpha^{3+} ),
\end{equation}
where
\begin{eqnarray}\label{Ys}
 Y^+_{\alpha}&=& a_1(\alpha^{1+}\cdot(p^2-p^3)+\alpha^{2+}\cdot(p^3-p^1)+\alpha^{3+}\cdot(p^1-p^2)), \nonumber \\
 Y^+_{gh.}&=&a_1(c^{1+}(b_0^3- b_0^2)+c^{2+}(b_0^1- b_0^3)+c^{3+}(b_0^2- b_0^1)),  \\ \nonumber
\end{eqnarray}
and
\begin{eqnarray}\label{Xs}
X^+_{\alpha}&=& 2a_3( (\alpha^{1+}\cdot\alpha^{2+})\alpha^{3+}\cdot(p^1-p^2)+
 (\alpha^{2+}\cdot\alpha^{3+})\alpha^{1+}\cdot(p^2-p^3)\nonumber\\
 &&+
 (\alpha^{3+}\cdot\alpha^{1+})\alpha^{2+}\cdot(p^3-p^1)),\\ \nonumber
 X^+_{gh.}&=&2a_3((\alpha^{1+}\cdot\alpha^{2+})c^{3+}(b_0^2-b_0^1)+
 (\alpha^{2+}\cdot\alpha^{3+})c^{1+}\cdot(b_0^3-b_0^2)\nonumber\\
 &&+
 (\alpha^{3+}\cdot\alpha^{1+})c^{2+}\cdot(b_0^1-b_0^3)).\nonumber
 \end{eqnarray}
As follows from the discussion above, BRST invariance (either on--shell or off--shell)  of the vertex (\ref{TOTAL}) is guaranteed by the fact that
each expression $\Delta_i$, $i=1,2,3$ is BRST invariant itself. If one considers only cubic interactions
one can take any functions of $\Delta_i$ and
obtain a valid cubic interaction vertex.
Since on the cubic level there are no further constraints on the vertices,
one has to investigate quartic interactions in order to understand what kind
of functional dependence is allowed in cubic vertices.

\setcounter{equation}0\section{Quartic vertices}\label{QV}

In this Section  we allow an arbitrary dependence of cubic vertices on the expressions $\Delta_1$
and $\Delta_3$ and we keep an exponential dependence on the expression $\Delta_2$.
The reason behind this is that in principle one can consider interaction vertices
for irreducible (Fronsdal) massless higher spin  modes rather than reducible ones.
Moreover, one can always decompose free Lagrangians which contain reducible modes into
Lagrangians which contain irreducible modes \cite{Fotopoulos:2009iw} and then build interactions among them.

Irreducible modes obey the on--shell constraint of being  traceless, therefore the Taylor expansion
of the function which contains $\Delta_2$ will terminate already at the linear term.
 Therefore, since functional dependence
of the cubic  vertex on $\Delta_2$ is not crucial, we shall consider for simplicity a cubic  vertex of the form
\begin{equation} \label{VERYTOTAL}
 V_3 |0\rangle_{123} = F(\Delta_1) Q({\Delta_3}) e^{\Delta_2}|0\rangle_{123},
\end{equation}
or equivalently
\begin{equation} \label{VERYTOTALTAYLOR}
 V_3 |0\rangle_{123} =\sum_{m,n,l=0}^\infty  \frac{F^{(m)}(\Delta_1)}{m!} \frac{Q^{(n)}({\Delta_3})}{n!} \frac{(\Delta_2)^l}{l!}|0\rangle_{123}.
\end{equation}
At the same time higher spin fields and parameters of gauge transformations are constrained
as in (\ref{EOM})--(\ref{gauget}) and, for the case of irreducible modes,
they obey zero trace constraint
\begin{equation} \label{0trace}
M^i |\phi_i \rangle= M^i |\lambda_i \rangle=0, \quad M^i = \frac{1}{2} \alpha^{i} \cdot \alpha^i,
\end{equation}
where no summation over $i$ is assumed.
Then the equation for finding on-shell quartic vertices is (\ref{quartons}).

Let us note that as a simple illustration of the method described in
the previous section one can take the spin of all interacting fields to
be equal to 1. Equation (\ref{Xs}) is then the colour-stripped Yang-Mills cubic
vertex and solving   (\ref{quartons}) for $| V_4 \rangle$ one obtains the
standard Yang-Mills four-vertex.

\subsection{Quartic vertex: $\Delta_1$ dependence}\label{QV2}

Let us consider first in detail
 the simplest example of the cubic vertex which contains only dependence on $\Delta_1$, i.e.
\begin{equation}\label{Delta1}
V_3|0\rangle_{123} = F\left( \Delta_1 \right)|0\rangle_{123}.
\end{equation}
Consider a term in the variation of the total Lagrangian, where the parameter of gauge transformations
is in the Hilbert space labelled by the index ``3". One has
\begin{eqnarray}
\delta_{1}|\phi_{1}\rangle & = & -g\int dc_{0}^{2}dc_{0}^{3}\langle\phi_{2}|\langle\Lambda_{3}||V_{3}(1,2,3)\rangle\nonumber \\
 & = & -g\int dc_{0}^{2}dc_{0}^{3}\langle\phi_{2}|\langle\lambda_{3}|b_{3}F\left(\Delta_{1}\right)c_{0}^{1}c_{0}^{2}c_{0}^{3}|0\rangle_{123}\nonumber \\
 & = & g\langle\phi_{2}|\langle\lambda_{3}|b_{3}\left[F(Y^+_{\alpha})+Y_{gh.}^+F^{(1)}(Y_\alpha^+)\right]c_{0}^{1}|0\rangle_{123}\nonumber \\
 & = & g\langle\phi_{2}|\langle\lambda_{3}|b_{3}Y_{gh.}^+F^{(1)}(Y_\alpha^{+})c_0^1|0\rangle_{123}
  =  -ga_{1}\langle\phi_{2}|\langle\lambda_{3}|F^{(1)}(Y_\alpha^{+})|0\rangle_{123}.
\end{eqnarray}
Let us consider the relevant expression in the variation of the Lagrangian which contains cubic vertices, i.e. the right hand side of
the equation (\ref{quartons}).
Initially we put the gauge parameter $\lambda$ in the Hilbert space labelled by the index $3^\prime$, and solve (\ref{quartons}). This corresponds to choosing just one of the possible permutations on the right hand side of \eqref{quartons}, and hence (after relabelling) one of the possible Hilbert spaces for the gauge parameter on the left hand side.
When considering equation (\ref{quartons}) we need to integrate over the first Hilbert space, which amounts to normal--ordering the $\alpha^1$ oscillators and using the fact that $\int dc_0^1\; c_0^1\;{}_1\langle 0||0\rangle_1=1$.
\begin{eqnarray*}
\delta\tilde{L} & = & -a_{1}g^{2}\int dc_{0}^{1}dc_{0}^{2}dc_{0}^{3}\langle V_{3}(1,2,3)||\phi_{2}\rangle|\phi_{3}\rangle\langle\phi_{2^{\prime}}|\langle\lambda_{3^{\prime}}|F^{(1)}(Y_\alpha^+)|0\rangle_{12^{\prime}3^{\prime}}\\
 & = & a_{1}g^{2}\,{}_{123}\langle0|F\left(Y_\alpha\right)|\phi_{2}\rangle|\phi_{3}\rangle\langle\phi_{2^{\prime}}
 |\langle\lambda_{3^{\prime}}|F^{(1)}\left(Y_\alpha^{+}\right)|0\rangle_{12^{\prime}3^{\prime}}\\
 & = & a_{1}g^{2}\,{}_{123}\langle0|\sum_{k=0}^{\infty}\frac{1}{k!}\left(a_{1}^{2}p_{23}\cdot p_{2^{\prime}3^{\prime}}\right)^{k}F^{(k)}\left[2a_{1}\left(\alpha^{2}\cdot p_{3}-\alpha^{3}\cdot p_{2}\right)\right]\\
 &  & \times|\phi_{2}\rangle|\phi_{3}\rangle\langle\phi_{2^{\prime}}|\langle\lambda_{3^{\prime}}|F^{(k+1)}\left[2a_{1}\left(\alpha^{2^{\prime}+}\cdot p_{3^{\prime}}-\alpha^{3^{\prime}+}\cdot p_{2^{\prime}}\right)\right]|0\rangle_{12^{\prime}3^{\prime}},
 \end{eqnarray*}
where we have used the notation $p_{\mu}^{ij}\equiv p^i_\mu -p^j_\mu$. Making use of (\ref{223}) and changing labels of the Hilbert
spaces as $2,3\rightarrow1,2$ and $2^{\prime},3^{\prime}\rightarrow3,4$,
one arrives at
\begin{eqnarray}
\delta\tilde{L} & = & a_{1}g^{2}\langle\phi_{1}|\langle\phi_{2}|\langle\phi_{3}|\langle\lambda_{4}|\sum_{k=0}^{\infty}\frac{1}{k!}\left(a_{1}^{2}p_{12}\cdot p_{34}\right)^{k}F^{(k)}\left[2a_{1}\left(\alpha^{1+}\cdot p_{2}-\alpha^{2+}\cdot p_{1}\right)\right]\nonumber \\
 &  & \times F^{(k+1)}\left[2a_{1}\left(\alpha^{3+}\cdot p_{4}-\alpha^{4+}\cdot p_{3}\right)\right]|0\rangle_{1234}.\label{eq:variation}
 \end{eqnarray}

Now, we take an ansatz for a quartic vertex
\begin{eqnarray}
|V_{4}(1,2,3,4)\rangle & = & G(p)\sum_{r,s,t=0}^{\infty}\frac{v_{rst}}{r!s!t!}\left(a_{1}^{2}p_{12}\cdot p_{34}\right)^{r}\left[2a_{1}\left(\alpha^{1+}\cdot p_{2}-\alpha^{2+}\cdot p_{1}\right)\right]^{s}\nonumber \\
 &  & \times\left[2a_{1}\left(\alpha^{3+}\cdot p_{4}-\alpha^{4+}\cdot p_{3}\right)\right]^{t}|0\rangle_{1234}.\end{eqnarray}
The relevant part of the left hand side of the equation (\ref{quartons}) is
\begin{eqnarray} \label{eq:QV4}
l^{4}V_{4} & = & -2a_{1}\left(p_{3}\cdot p_{4}\right)G(p)\sum_{r,s,t=0}^{\infty}\frac{v_{r,s,t+1}}{r!s!t!}\left(a_{1}^{2}p_{12}\cdot p_{34}\right)^{r}
\\
&  & \times \left[2a_{1}\left(\alpha^{1+}\cdot p_{2}-\alpha^{2+}\cdot p_{1}\right)\right]^{s}
  \left[2a_{1}\left(\alpha^{3+}\cdot p_{4}-\alpha^{4+}\cdot p_{3}\right)\right]^{t} \nonumber .
 \end{eqnarray}
 Now, equating \eqref{eq:QV4} and \eqref{eq:variation}, one has
\begin{eqnarray}\label{G1}
G(p) & = & -\frac{1}{s},\\
v_{k,m,n+1} & = & F^{(k+m)}(0)F^{(k+n+1)}(0), \nonumber
\end{eqnarray}
where we use the standard
definitions for Mandelstam variables $s=(p_1+p_2)^2$,
$t=(p_1+p_4)^2$ and $u=(p_1+p_3)^2$.
 The second equation implies that $v_{kmn}=F^{(k+m)}(0)F^{(k+n)}(0)$, and
finally one arrives to the expression for the quartic vertex
\begin{eqnarray} \label{eq:quarticvertex1}
|V_{4}\rangle_s & = & -\frac{1}{s}\sum_{k,m,n=0}^{\infty}\frac{F^{(k+m)}(0)F^{(k+n)}(0)}{k!m!n!}\left(a_{1}^{2}p_{12}\cdot p_{34}\right)^{k}
 \\
&  & \times \left[2a_{1}\left(\alpha^{1+}\cdot p_{2}-\alpha^{2+}\cdot p_{1}\right)\right]^{m}
 \left[2a_{1}\left(\alpha^{3+}\cdot p_{4}-\alpha^{4+}\cdot p_{3}\right)\right]^{n}|0\rangle_{1234} \nonumber.
 \end{eqnarray}

In the equation above, the notation $|V_4\rangle_s$ means that this part of the quartic vertex has a pole in the s-channel. The full solution is given by acting with the vertex (\ref{eq:quarticvertex1})
 on all non-cyclic permutations of the external states
\begin{equation}\label{LV4}
 \langle 1,2,3,4| V_4\rangle_s + \langle 1,3,2,4|
V_4\rangle_u + \langle 1,4,2,3| V_4\rangle_t, \end{equation} where
each contribution has subscript indicating the massless pole on
the corresponding kinematic variable which comes from the
definition of $G(p)$ in (\ref{G1}).

Taking, for example, the function $F(Y_\alpha^+)$ to be an exponential
one recovers an on--shell version of the off--shell quartic vertex
considered in \cite{Tsulaia:2012rb}, \cite{Fotopoulos:2010ay}.
One can also immediately write an off--shell
version of the quartic vertices obtained in this section.
Indeed, repeating the calculations of the present subsection and of
\cite{Fotopoulos:2010ay} one arrives at the following expression
for an off--shell quartic vertex
\begin{eqnarray} \label{eq:quarticvertex1off} \nonumber
|V_{4}\rangle_s & = & -\frac{1}{s}\sum_{k,m,n=0}^{\infty}\frac{F^{(k+m)}(0)F^{(k+n)}(0)}{k!m!n!}\left(a_{1}^{2}p_{12}\cdot p_{34}\right)^{k}
 \\
& & \times \left[a_{1}\left(2\alpha^{1+}\cdot p_{2}-2\alpha^{2+}\cdot p_{1} - c^{1+}b_0^2+ c^{2+}b_0^1\right)\right]^{m} \\ \nonumber
& & \times \left[a_{1}\left(2\alpha^{3+}\cdot p_{4}-2\alpha^{4+}\cdot p_{3}  - c^{3+}b_0^4+ c^{4+}b_0^3  \right)\right]^{n}|0\rangle_{1234},
\end{eqnarray}
which solves the off--shell equation (\ref{quartbrst}) for quartic vertices.

\subsection{Quartic vertex:  $\Delta_1$ and $\Delta_2$ dependence}\label{QV3}

We now consider the
case where
\begin{equation}\label{Delta2}
V_{3}|0\rangle_{123} =F\left(\Delta_1\right)\exp\left(\Delta_2\right)|0\rangle_{123}.
\end{equation}
Repeating the steps of the previous subsection (see the Appendix for the technical details)
we take as an ansatz for the associated quartic vertex
\begin{eqnarray}\label{v42}
V_4 &=& G(p)\sum_{m,n,r,s,t=0}^\infty \frac{v_{mnrst}}{m!n!r!s!t!}\left(-e^\beta\frac{a_1^2 a_2}{2}s\right)^{m+n}\left(e^\beta a_1^2 p_{12}\cdot p_{34}\right)^{t} \nonumber \\
  & &\times \left(2a_1\left(\alpha^{1+}\cdot p_2-\alpha^{2+}\cdot p_1\right)\right)^r
  \left(2a_1\left(\alpha^{3+}\cdot p_{4}-\alpha^{4+}\cdot p_{3}\right)\right)^s\nonumber \\
  & &\times \exp\left(\frac{D}{2}\beta +a_2\left(M^{1+}+M^{2+}+M^{3+}+M^{4+}\right)\right), \nonumber \\
\end{eqnarray}
where $\beta=-\log(1-a_2^2)$.
Inserting (\ref{v42}) into \eqref{quartons} one can see that the coefficients $v_{mnrst}$ and the function $G(p)$ are determined to be
\begin{eqnarray}\label{G2}
G(p) &=& -\frac{1}{s}, \\
v_{mnrst} &=& F^{(2m+r+t)}\!(0)F^{(2n+s+t)}\!(0). \nonumber
\end{eqnarray}
Finally, we find that the quartic vertex is
\begin{eqnarray}\label{eq:quarticvertex2}
|V_4\rangle_s &=& -\frac{1}{s}\sum_{m,n,r,s,t=0}^\infty \frac{F^{(2m+r+t)}\!(0)F^{(2n+s+t)}\!(0)}{m!n!r!s!t!} \nonumber \\
& & \times \left(-e^\beta\frac{a_1^2 a_2}{2}s\right)^{m+n}\left(e^\beta a_1^2 p_{12}\cdot p_{34}\right)^{t} \nonumber \\
& &\times
  \left(2a_1\left(\alpha^{1+}\cdot p_2-\alpha^{2+}\cdot p_1\right)\right)^r \left(2a_1\left(\alpha^{3+}\cdot p_{4}-\alpha^{4+}\cdot p_{3}\right)\right)^s \nonumber \\
  & &\times \exp\left(\frac{D}{2}\beta +a_2\left(M^{1+}+M^{2+}+M^{3+}+M^{4+}\right)\right) |0\rangle_{1234}. \nonumber \\
\end{eqnarray}

\subsection{Quartic vertex:  $\Delta_1$ and $\Delta_3$ dependence}\label{QV4}

Let us take the model which describes irreducible massless higher spin modes, i.e.
when the higher spin fields are traceless on-shell. Taking
\begin{equation}\label{1-3}
V_3|0\rangle_{123}=F\left(\Delta_1\right)Q\left(\Delta_3\right)|0\rangle_{123},
\end{equation}
and repeating the steps of previous subsections,
we consider a term in the variation of the total Lagrangian where the gauge parameter is in the Hilbert space labelled by ``3''. One has
\begin{eqnarray}
\delta_1|\phi_1\rangle &=& -g\int dc_{0}^{2}dc_{0}^{3}\langle\phi_{2}|\langle\lambda_{3}|b_3|V_{3}(1,2,3)\rangle\nonumber \\
&=& g\langle\phi_{2}|\langle\lambda_{3}|\left[F\left(Y_{\alpha}^+\right) \tilde{X}^{(3)}_{rs31}\left(\alpha^{r+}\cdot\alpha^{s+}\right) Q^{\prime}\left(X_{\alpha}^+\right)\right. \nonumber \\
& &\hspace{20mm}+\left.F^{\prime}\left(Y_{\alpha}^+\right)Z_{31}Q\left(X_{\alpha}^+\right)\right] |0\rangle_{123} \nonumber \\
&=& -g\langle\phi_{2}|\langle\lambda_{3}|\left[2a_3 F\left(Y_{\alpha}^+\right) \left(\alpha^{1+}\cdot\alpha^{2+}\right) Q^{\prime}\left(X_{\alpha}^+\right) \right. \nonumber \\
& &\hspace{25mm}\left. +a_1 F^{\prime}\left(Y_{\alpha}^+\right)Q\left(X_{\alpha}^+\right)\right] |0\rangle_{123}. \nonumber \\
\end{eqnarray}
From this nonlinear gauge transformation we find
\begin{eqnarray*}
\delta\tilde{L} &=& -g^{2}\int dc_{0}^{1}dc_{0}^{2}dc_{0}^{3}\langle V_{3}(1,2,3)||\phi_{2}\rangle|\phi_{3}\rangle \langle\phi_{2^{\prime}}|\langle\lambda_{3^{\prime}}| \\
& &\times
\left[2a_3 F\left(Y_{\alpha}^+\right) \left(\alpha^{1+}\cdot\alpha^{2^{\prime}+}\right) Q^{(1)}\left(X_{\alpha}^+\right)
+a_1 F^{(1)}\left(Y_{\alpha}^+\right)Q\left(X_{\alpha}^+\right)\right]|0\rangle_{12^{\prime}3^{\prime}} \\
&=& g^2\,{}_{123}\langle 0|F\left(Y_\alpha\right)Q\left(X_\alpha\right)|\phi_{2}\rangle|\phi_{3}\rangle \langle\phi_{2^{\prime}}|\langle\lambda_{3^{\prime}}| \\
& & \times
\left[2a_3 F\left(Y_{\alpha}^+\right) \left(\alpha^{1+}\cdot\alpha^{2^{\prime}+}\right) Q^{(1)}\left(X_{\alpha}^+\right)
+a_1 F^{(1)}\left(Y_{\alpha}^+\right)Q\left(X_{\alpha}^+\right)\right]|0\rangle_{12^{\prime}3^{\prime}}. \\
\end{eqnarray*}
Performing the commutators to integrate out the ``1'' Hilbert space, and changing labels  of the
 remaining Hilbert spaces as in \eqref{eq:variation}, one arrives at
\begin{eqnarray}
\delta\tilde{L} &=& g^2\langle\phi_{1}|\langle\phi_{2}|\langle\phi_{3}|\langle\lambda_{4}|
\sum_{k,l,m,n,p,q=0}^{\infty}\frac{1}{k!l!m!n!p!q!} \nonumber \\
& &\times \left[2a_1 a_3\left(\left(\alpha^{3+}\cdot\alpha^{4+}\right)p_{12}\cdot p_{34} +2\left(\alpha^{4+}\cdot p_{12}\right)\left(\alpha^{3+}\cdot p_4\right)\right.\right. \nonumber \\
& &\hspace{75mm}\left.\left.-2\left(\alpha^{3+}\cdot p_{12}\right)\left(\alpha^{4+}\cdot p_3\right)\right)\right]^k \nonumber \\
& &\times \left[2a_1 a_3\left(\left(\alpha^{1+}\cdot\alpha^{2+}\right)p_{12}\cdot p_{34} +2\left(\alpha^{2+}\cdot p_{34}\right)\left(\alpha^{1+}\cdot p_2\right)\right.\right. \nonumber \\
& &\hspace{75mm}\left.\left.-2\left(\alpha^{1+}\cdot p_{34}\right)\left(\alpha^{2+}\cdot p_1\right)\right)\right]^l \nonumber \\
& &\times \left[a_1^2\, p_{12}\cdot p_{34}\right]^m \left[2a_1\left(\alpha^{1+}\cdot p_2-\alpha^{2+}\cdot p_1\right)\right]^p \left[2a_1\left(\alpha^{3+}\cdot p_4-\alpha^{4+}\cdot p_3\right)\right]^q \nonumber \\
& &\times \left[4a_3^2\left(4\left(\alpha^{2+}\cdot p_1\right)\left[\left(\alpha^{1+}\cdot\alpha^{3+}\right)\left(\alpha^{4+}\cdot p_3\right)-\left(\alpha^{1+}\cdot\alpha^{4+}\right)\left(\alpha^{3+}\cdot p_4\right)\right]\right.\right. \nonumber \\
& & \quad -4\left(\alpha^{1+}\cdot p_2\right)\left[\left(\alpha^{2+}\cdot\alpha^{3+}\right)\left(\alpha^{4+}\cdot p_3\right)-\left(\alpha^{2+}\cdot\alpha^{4+}\right)\left(\alpha^{3+}\cdot p_4\right)
\right] \nonumber \\
& & \quad +2\left(\alpha^{1+}\cdot\alpha^{2+}\right)\left[\left(\alpha^{4+}\cdot p_{12}\right)\left(\alpha^{3+}\cdot p_4\right)
-\left(\alpha^{3+}\cdot p_{12}\right)\left(\alpha^{4+}\cdot p_3\right)\right] \nonumber \\
& & \quad +2\left(\alpha^{3+}\cdot\alpha^{4+}\right)\left[\left(\alpha^{2+}\cdot p_{34}\right)\left(\alpha^{1+}\cdot p_2\right)
-\left(\alpha^{1+}\cdot p_{34}\right)\left(\alpha^{2+}\cdot p_1\right)\right] \nonumber \\ 
& &\left.\left.  +\left(\alpha^{1+}\cdot\alpha^{2+}\right)\left(\alpha^{3+} \cdot\alpha^{4+}\right)p_{12}\cdot p_{34} \right)\right]^n
 \nonumber \\
& &\times\left\{a_1 F^{(m+k+p)}(0)F^{(m+l+q+1)}(0)Q^{(l+n)}(0)Q^{(k+n)}(0)\right. \nonumber \\
& & \hspace{5mm} +2a_1a_3\left(\alpha^{3+}\cdot p_{12}\right)F^{(m+k+p+1)}(0)F^{(m+l+q)}(0)Q^{(l+n)}(0)Q^{(k+n+1)}(0) \nonumber \\
& & \hspace{5mm} +4a_3^2\left[\left(\alpha^{1+}\cdot\alpha^{2+}\right)\alpha^{3+}\cdot p_{12} + 2\left(\alpha^{2+}\cdot\alpha^{3+}\right)\alpha^{1+}\cdot p_{2}\right. \nonumber \\
& &\hspace{75mm}\left.-2\left(\alpha^{1+}\cdot\alpha^{3+}\right)\alpha^{2+}\cdot p_{1}\right] \nonumber \\
& &\quad\left. \times F^{(m+k+p)}(0)F^{(m+l+q)}(0)Q^{(l+n+1)}(0)Q^{(k+n+1)}(0)\right\}|0\rangle_{1234}. \label{deltaL13}
\end{eqnarray}
Now we take as a quartic ansatz
\begin{eqnarray}
V_4 &=& G(p)\sum_{k,l,m,n,p,q=0}^{\infty}\frac{v_{klmnpq}}{k!l!m!n!p!q!} \nonumber \\
& &\times \left[2a_1 a_3\left(\left(\alpha^{3+}\cdot\alpha^{4+}\right)p_{12}\cdot p_{34} +2\left(\alpha^{4+}\cdot p_{12}\right)\left(\alpha^{3+}\cdot p_4\right)\right.\right. \nonumber \\
& &\hspace{70mm}\left.\left.-2\left(\alpha^{3+}\cdot p_{12}\right)\left(\alpha^{4+}\cdot p_3\right)\right)\right]^k \nonumber \\
& &\times \left[2a_1 a_3\left(\left(\alpha^{1+}\cdot\alpha^{2+}\right)p_{12}\cdot p_{34} +2\left(\alpha^{2+}\cdot p_{34}\right)\left(\alpha^{1+}\cdot p_2\right)\right.\right. \nonumber \\
& &\hspace{70mm}\left.\left.-2\left(\alpha^{1+}\cdot p_{34}\right)\left(\alpha^{2+}\cdot p_1\right)\right)\right]^l \nonumber \\
& &\times \left[a_1^2\, p_{12}\cdot p_{34}\right]^m \left[2a_1\left(\alpha^{1+}\cdot p_2-\alpha^{2+}\cdot p_1\right)\right]^p \left[2a_1\left(\alpha^{3+}\cdot p_4-\alpha^{4+}\cdot p_3\right)\right]^q \nonumber \\
& &\times \left[4a_3^2\left(4\left(\alpha^{2+}\cdot p_1\right)\left[\left(\alpha^{1+}\cdot\alpha^{3+}\right)\left(\alpha^{4+}\cdot p_3\right)-\left(\alpha^{1+}\cdot\alpha^{4+}\right)\left(\alpha^{3+}\cdot p_4\right)\right]\right.\right. \nonumber \\
& & \quad -4\left(\alpha^{1+}\cdot p_2\right)\left[\left(\alpha^{2+}\cdot\alpha^{3+}\right)\left(\alpha^{4+}\cdot p_3\right)-\left(\alpha^{2+}\cdot\alpha^{4+}\right)\left(\alpha^{3+}\cdot p_4\right)
\right] \nonumber \\
& & \quad +2\left(\alpha^{1+}\cdot\alpha^{2+}\right)\left[\left(\alpha^{4+}\cdot p_{12}\right)\left(\alpha^{3+}\cdot p_4\right)
-\left(\alpha^{3+}\cdot p_{12}\right)\left(\alpha^{4+}\cdot p_3\right)\right] \nonumber \\
& & \quad +2\left(\alpha^{3+}\cdot\alpha^{4+}\right)\left[\left(\alpha^{2+}\cdot p_{34}\right)\left(\alpha^{1+}\cdot p_2\right)
-\left(\alpha^{1+}\cdot p_{34}\right)\left(\alpha^{2+}\cdot p_1\right)\right] \nonumber \\
& & \quad \left.\left.  +\left(\alpha^{1+}\cdot\alpha^{2+}\right)\left(\alpha^{3+} \cdot\alpha^{4+}\right)p_{12}\cdot p_{34} \right)\right]^n. \nonumber \\
\end{eqnarray}
The relevant part of the left hand side of equation (\ref{quartons}) is $l^4V_4$, which upon equating with the expression \eqref{deltaL13} gives
\begin{eqnarray} \label{G3}
G(p) &=& -\frac{1}{s}, \\
v_{klmnpq} &=& F^{(m+k+p)}(0)F^{(m+l+q)}(0)Q^{(l+n)}(0)Q^{(k+n)}(0). \nonumber
\end{eqnarray}
Therefore, the expression for the quartic vertex is
\begin{eqnarray}\label{eq:quarticvertex3}
|V_4\rangle_s &=& -\frac{1}{s}\sum_{k,l,m,n,p,q=0}^{\infty}\frac{F^{(m+k+p)}(0)F^{(m+l+q)}(0)Q^{(l+n)}(0)Q^{(k+n)}(0)}{k!l!m!n!p!q!} \nonumber \\
& &\times \left[2a_1 a_3\left(\left(\alpha^{3+}\cdot\alpha^{4+}\right)p_{12}\cdot p_{34} +2\left(\alpha^{4+}\cdot p_{12}\right)\left(\alpha^{3+}\cdot p_4\right)\right.\right. \nonumber \\
& &\hspace{60mm}\left.\left.-2\left(\alpha^{3+}\cdot p_{12}\right)\left(\alpha^{4+}\cdot p_3\right)\right)\right]^k \nonumber \\
& & \times\left[2a_1 a_3\left(\left(\alpha^{1+}\cdot\alpha^{2+}\right)p_{12}\cdot p_{34} +2\left(\alpha^{2+}\cdot p_{34}\right)\left(\alpha^{1+}\cdot p_2\right)\right.\right. \nonumber \\
& &\hspace{60mm} \left.\left.-2\left(\alpha^{1+}\cdot p_{34}\right)\left(\alpha^{2+}\cdot p_1\right)\right)\right]^l \nonumber \\
& &\times\left[a_1^2\, p_{12}\cdot p_{34}\right]^m \left[2a_1\left(\alpha^{1+}\cdot p_2-\alpha^{2+}\cdot p_1\right)\right]^p \left[2a_1\left(\alpha^{3+}\cdot p_4-\alpha^{4+}\cdot p_3\right)\right]^q \nonumber \\
& & \times \left[4a_3^2\left(4\left(\alpha^{2+}\cdot p_1\right)\left[\left(\alpha^{1+}\cdot\alpha^{3+}\right)\left(\alpha^{4+}\cdot p_3\right)-\left(\alpha^{1+}\cdot\alpha^{4+}\right)\left(\alpha^{3+}\cdot p_4\right)\right]\right.\right. \nonumber \\
& & \quad -4\left(\alpha^{1+}\cdot p_2\right)\left[\left(\alpha^{2+}\cdot\alpha^{3+}\right)\left(\alpha^{4+}\cdot p_3\right)-\left(\alpha^{2+}\cdot\alpha^{4+}\right)\left(\alpha^{3+}\cdot p_4\right)
\right] \nonumber \\
& & \quad +2\left(\alpha^{1+}\cdot\alpha^{2+}\right)\left[\left(\alpha^{4+}\cdot p_{12}\right)\left(\alpha^{3+}\cdot p_4\right)
-\left(\alpha^{3+}\cdot p_{12}\right)\left(\alpha^{4+}\cdot p_3\right)\right] \nonumber \\
& & \quad +2\left(\alpha^{3+}\cdot\alpha^{4+}\right)\left[\left(\alpha^{2+}\cdot p_{34}\right)\left(\alpha^{1+}\cdot p_2\right) \right. \nonumber \\
& & \quad \left.\left.\left.-\left(\alpha^{1+}\cdot p_{34}\right)\left(\alpha^{2+}\cdot p_1\right)\right]
+\left(\alpha^{1+}\cdot\alpha^{2+}\right)\left(\alpha^{3+} \cdot\alpha^{4+}\right)p_{12}\cdot p_{34}
\right)\right]^n|0\rangle_{1234}. \nonumber \\
\end{eqnarray}
Taking the functions $F$ and $Q$ to be exponentials and choosing
$a_1= 2a_3 = \sqrt{\frac{\alpha^\prime}{2}}$ one recovers a quartic vertex obtained in
\cite{Sagnotti:2010at}.

In all examples considered in this Section the
full solution is given by acting with the vertices (\ref{eq:quarticvertex1}),
(\ref{eq:quarticvertex2}), (\ref{eq:quarticvertex3})
 on all non-cyclic permutations of the external states, where
each contribution has subscript indicating the massless pole on
the corresponding kinematic variable which comes from the
definition of $G(p)$ in (\ref{G1}), (\ref{G2}), and (\ref{G3}).

\setcounter{equation}0\section{Conditions on the S-matrix}\label{SM}

As we have seen in the examples considered in the previous Section
the procedure of computing quartic vertices from the known cubic
ones does not impose any restriction on the unknown functions which are present in the definition
of cubic vertices.
However, on a flat background there can be extra constraints
on cubic, and therefore on quartic, vertices which come from
the requirement of the existence of a nontrivial $S$--matrix.
Below we briefly recall these requirements
and then apply them to the theory under consideration.
We shall mainly follow \cite{Fotopoulos:2010ay}
where a more detailed discussion can be found.

The consideration is based on the BCFW recursion relations
\cite{Britto:2004ap}--\cite{Britto:2005fq}
and the consistency test introduced in \cite{Benincasa:2007xk}.
The key point of the BCFW method is that tree level
amplitudes constructed using Feynman rules are rational functions
of external momenta. Analytic continuation of these momenta on the
complex domain  turns the amplitudes into meromorphic functions
which can be constructed solely by their residues. Since the
residues of scattering amplitudes are, due to unitarity, products
of lower--point on--shell amplitudes the final outcome is a set of
powerful recursion relations.

The simplest complex deformation (BCFW shift) involves only two external
particles whose momenta are shifted as
\begin{eqnarray}\label{BCFWshift1}
 \hp_i(z)= p_i - q z\ , \quad
 \hp_j(z)= p_j + q z \ .
\end{eqnarray}
Here $z \in \mathbb{C}$ and, to keep the on-shell condition, we need
$q \cdot p_{i} =q\cdot p_j= 0$ and  $q^2 = 0$. In Minkowski
space--time this is only possible for complex $q$.
In
four dimensions we can use a spinor representation of momenta and
polarizations (for spin 1 states)
\begin{eqnarray}\label{spinor}
&&p^\mu = \lambda^a (\sigma^\mu)_{a\dot a}\tilde\lambda^{\dot
a},\nonumber \\
 &&\epsilon^+_{a\dot a} = \frac{\mu_a\tilde\lambda_{\dot
a}}{\langle \mu ,\lambda \rangle},  \qquad  \epsilon^-_{a\dot a}
=\frac{\lambda_a\tilde\mu_{\dot a}}{[\tilde\lambda , \tilde\mu]}, \\
&& \langle \mu ,\lambda \rangle \equiv \mu_a \lambda_b
\epsilon^{ab}, \qquad [\tilde\lambda , \tilde\mu] \equiv
\tilde\mu_{\dot a} \tilde\lambda_{\dot b} \epsilon^{\dot a \dot
b},\nonumber
\end{eqnarray}
with $\mu_a$ and $\tilde\mu_{\dot a}$ arbitrary reference spinors.
Polarizations of higher spin states  are given by products of the
polarizations for spin 1
\begin{equation}\label{polten}
\epsilon^+_{a_1 \dot{a}_1 \dots a_s \dot{a}_s} = \prod_{i=1}^s
\epsilon ^+_{a_i \dot{a}_i}, \qquad \epsilon^-_{a_1 \dot{a}_1 \dots
a_s \dot{a}_s} = \prod_{i=1}^s \epsilon ^-_{a_i \dot{a}_i}.
\end{equation}
The undeformed amplitude
can be computed using Cauchy's theorem:
\begin{equation}\label{BCFWrel1}
{\cal M}_n (0) = \frac{1}{2\pi i}\oint_{z=0} \frac{{\cal M}_n(z)}{z} dz = -
\left\{\sum \mathrm{Res}_{z=\textrm{finite}} + \mathrm{Res}_{z=
\infty} \right\} \ .
\end{equation}
Whereas the
 residues at finite locations on the complex
plane are  products of lower--point
tree level amplitudes which are computed at complex on--shell momenta,
the residue at  infinity in general does not have such an interpretation.
The theories where the deformed amplitude vanishes at complex infinity
are called constructable theories \cite{Benincasa:2007xk}.
In such  theories all four point amplitudes are expressed in terms of three point ones.
In \cite{Benincasa:2007xk} has been derived the
criterion of consistency of four point functions
which is also a
necessary condition for a theory to have zero residue at infinity.
Let us denote by ${\cal
M}^{(i,j)}(z)$ the four-point function under deformation of
particles $i$ and $j$. Assume further  that the helicities $h_2$
and $h_4$ are negative while $h_1$ is positive.  The criterion
advocates that
\begin{equation}\label{BC1}
{\cal M}_4^{(1,2)}(0)={\cal M}_4^{(1,4)}(0).
\end{equation}
This is highly non-trivial since the usual Feynman analysis
construction of the four-point function requires adding diagrams
from three possible channels, while each BCFW deformation can use
only two channels of exchanged particles. For ${\cal M}^{(1,2)}(0)$,
diagrams where particles 1 and 2 go to an intermediate state do
not lead to poles in the complex plane, i.e. $1/
(p_1(z)+p_2(z))^2=1/ (p_1+p_2)^2$. So in ${\cal M}^{(1,2)}(z)$ only poles
from the $t,\,u$ channels in the complex $z$-plane will contribute.
Similarly, for ${\cal M}^{(1,4)}(0)$ only poles from the $s,\,u $
channels in the complex $z$-plane will contribute.

Let us first recall that, for any Lorentz invariant theory of massless particles in
four dimensions,  on-shell cubic
amplitudes vanish. Considering complex momenta
one  can derive that the most generic cubic
amplitude for particles with helicities $h_1$, $h_2$ and $h_3$ is given by \cite{Benincasa:2007xk}
\begin{equation}\label{cubicamp}
{\cal M}_3(\{\lambda^{(i)},\tilde\lambda^{(i)},h_i\}) = \kappa_H
\langle 1,2\rangle^{d_{3}}\langle 2,3\rangle^{d_{1}}\langle
3,1\rangle^{d_{2}}+ \kappa_A [1,2]^{-d_{3}}[ 2,3]^{-d_{1}}[
3,1]^{-d_{2}},
\end{equation}
where  $d_1=h_1-h_2-h_3, \ d_2=h_2-h_1-h_3, \
d_3=h_3-h_1-h_2$ and the coupling constant $\kappa_H$ ($\kappa_A$) is
required to give the right dimension to the part of the amplitude
which is holomorphic (antiholomorphic) in the spinor variables
$\lambda^{(i)}$ ($\tilde{\lambda}^{(i)}$). Further,
 an explicit expression for the four--point function  ${\cal M}_4^{(1,2)}(0)$ is
\cite{Benincasa:2007xk}
\begin{eqnarray}\label{M12}
{\cal M}_4^{(1,2)}(0) = && \sum_{h > {\rm
max}(-(h_1+h_4),(h_2+h_3))} \left(
\kappa^A_{1-h_1-h_4-h}\kappa^H_{1+h_2+h_3-h}
\frac{(-P_{3,4}^2)^h}{P_{1,4}^2} \right.
 \nonumber \\
&& \hspace{10mm}\times\left(\frac{[1,4][3,4]}{[1,3]}\right)^{h_4}
\left(\frac{[1,3][1,4]}{[3,4]}\right)^{h_1} \nonumber \\
&& \hspace{30mm}\left.\times\left(\frac{\langle3,4\rangle}{\langle 2,3\rangle\langle2,4\rangle}\right)^{h_2}
\left(\frac{\langle 2,4\rangle}{\langle2,3\rangle\langle 3,4\rangle}\right)^{h_3}\right)  \\ \nonumber
&&+ \sum_{h
> {\rm
max}(-(h_1+h_3),(h_2+h_4))}\!\!\!\!\!\!\!\!(4\leftrightarrow
3),\nonumber
\end{eqnarray}
where $P_{i,j}= p_i +p_j$ and
the subscript of the coupling
constants denotes their mass dimension. The expression for
${\cal M}_4^{(1,4)}(0)$ can be simply obtained by exchanging labels $2$ and $4$
in (\ref{M12}).

After all these preliminaries we consider a simple example of the scattering of four real scalar fields
where the intermediate propagator is an infinite tower of irreducible massless higher spin fields,
similar to what has been done in \cite{Fotopoulos:2010ay}.
The cubic interaction
between two scalars and a massless higher spin field $\Psi_h^{\mu_1\dots \mu_h}$ has the form
 \cite{Berends:1985xx}
\begin{equation}\label{ILW}
{\cal L}_{int}^{00s}=   \kappa^{1-h}  N_{h} \ {\Psi_h^{\mu_1\dots \mu_h}
J^{1;2}_{h; \mu_1\dots \mu_h}} + \ h.c.\,,
\end{equation}
where the combination $\kappa N_h$
plays the role of a dimensionful coupling constant, and
 the currents are given by
\begin{equation}\label{Jp}
J^{1;2}_{h;\mu_1\dots \mu_h}=\sum_{r=0}^{h} \ \bn{h}{r} \ (-1)^r \
(\partial^{\mu_1} \dots
\partial^{\mu_r}
\phi_1)\ (\partial^{\mu_{r+1}} \dots
\partial^{\mu_{h}}\phi_2).
\end{equation}
Let us note that
since we are considering real scalar fields the spin of the intermediate higher spin field should be even.

As follows from the discussion above, an explicit dependence of the coupling constant
on the spin of the intermediate field is crucial for checking the relation (\ref{BC1}).
In other words one needs an explicit expression for $N_h$.
This can be easily established from (\ref{VERYTOTAL}) to
be\footnote{The simplest way to see it as follows:
the kinetic term for a higher spin field $A$ has the form
$L_k = \langle \phi | Q | \phi \rangle \sim \frac{p^2 A^2}{h!}$,
while the cubic interaction has the form  $V= \langle 0 |\phi^2 \frac{A \alpha^h}{h!}   \frac{(p \alpha^+)^h}{h!} F^{(h)}(0)|0 \rangle
\sim \phi^2 A \frac{F^{(h)}(0)}{h!}$. In order to bring the kinetic term to the canonical form
 one redefines the field $\Psi = \frac{A}{\sqrt{h!}}$
and  gets
$L= p^2 \Psi^2 + \frac{F^{(h)}(0)}{\sqrt {h!}}  p^h\phi^2 \Psi$.}
$\frac{F^{(h)}(0)}{\sqrt{h!}}$.
Finally, using (\ref{M12}) and
\be
\kappa^A_{1-h}=\kappa^H_{1-h}=\kappa^{1-h} N_h, \quad N_h=\frac{F^{(h)}(0)}{\sqrt{h!}},
\ee
one gets
\be \label{12}
{\cal M}_4^{(1,2)}(0) =\sum_{h \in 2\mathbb{N}} k^{2-2h}\frac{(F^{(h)}(0))^2}{h!}s^h (\frac{1}{t} + \frac{1}{u})
= \sum_{h \in 2\mathbb{N}} k^{2-2h}\frac{(F^{(h)}(0))^2}{h!}s^h(-\frac{s}{tu}),
\ee
and
\be \label{14}
{\cal M}_4^{(1,4)}(0) =\sum_{h \in 2\mathbb{N}} k^{2-2h}\frac{(F^{(h)}(0))^2}{h!}t^h (\frac{1}{u} + \frac{1}{s})
= \sum_{h \in 2\mathbb{N}} k^{2-2h}\frac{(F^{(h)}(0))^2}{h!}t^h(-\frac{t}{su}).
\ee
 From (\ref{12}) and  (\ref{14}) one can see that the equation can only be satisfied
 if the function F is zero due to the different
 functional dependence on the Mandelstam variables $s$ and $t$  in
 ${\cal M}_4^{(1,2)}(0)$ and ${\cal M}_4^{(1,4)}(0)$.
 This result could have been anticipated from a discussion
 given in \cite{Fotopoulos:2010ay}, according to which the
 massless higher spin theories on Minkowski space need some extra ingredients
 for their consistency, like nonlocal or composite objects.
 The requirement for these kind of objects is not obvious from the discussion above because
 we considered a general functional dependence in the cubic vertices (\ref{VERYTOTAL}),
 whereas in \cite{Fotopoulos:2010ay} explicit examples
 (both in spinorial formalisms and equivalently  in terms of standard Feynman diagrams)
  of $N_h=\textrm{const.}$ and $N_h= \frac{1}{\sqrt{h!}}$
  have been considered.  One can conclude from the explicit calculations
 presented above that, if one allows for general functions in
 the cubic vertices (\ref{VERYTOTAL}),  computation of the quartic
 vertices does not restrict their form. On the
 other hand the analysis of symmetries of the $S$-matrix, with the help
 of criteria proposed in \cite{Benincasa:2007xk}, indicates the triviality
 of higher order interactions.

 It is important to stress that the aforementioned
conclusion  about the $S$--matrix which is based upon the four particle test
(\ref{BC1}) is valid when the theory is constructable, i.e., when the total four--point amplitude
vanishes at complex infinity after the BCFW shift (\ref{BCFWshift1}). If the theory is not constructable
then the four particle test gives no conclusion about the properties of the $S$--matrix.
One can of course check directly under which conditions on the functions  $P$ and $Q$
in (\ref{VERYTOTAL})
the theory determined by the cubic vertices described in the  previous Section
is constructable. However our approach is slightly different. We assumed that the theory was constructable and
imposed the four particle test on the four--point functions to find that for constructable theories this test cannot be passed.

\setcounter{equation}0\section{AdS Deformation}\label{QADS}

It is well known that ${\cal D}$--dimensional
anti-de Sitter space is a natural background for
consistent interactions among massless higher spin fields \cite{Vasiliev:1990en}.
For this reason it would be interesting to generalize the discussion of the previous Sections
for the case of $AdS_{\cal D}$ background.
Cubic vertices \cite{Vasiliev:2001wa}--\cite{Lee:2012ku} as well as current--current interactions on $AdS_{\cal D}$ \cite{Francia:2008hd}
have been extensively discussed recently.
  In this section we describe how one can perform analogous computations on
anti-de Sitter space for quartic vertices. We mainly outline the procedure of finding
cubic and quartic vertices and give the important identities needed
for these calculations. We will leave detailed analysis for a future publication.

In the case of $AdS_{\cal D}$ space \cite{Buchbinder:2006ge}, the operators $p_\mu$ become
\begin{equation}
\label{pop}
p_\mu \ = \  -\; i \, \left(
\nabla_\mu + \omega_{\mu}^{ab} \, \alpha_{\; a}^+\,
  \alpha_{ \; b} \right),
\end{equation}
where    $\omega_\mu^{ab}$ is  the spin
connection and $\nabla_\mu$  the covariant derivative of $AdS_{\cal D}$.
Covariant derivatives
no longer commute, rather they satisfy the commutation relations
\begin{equation}
   \label{COMU}
   D_{\mu \nu} \equiv [p_\mu,p_\nu]
= -[\nabla_\mu,\nabla_\nu]+ \frac{1}{L^2}\; (\alpha_{\; \mu}^+ \, \alpha_{\; \nu} \, -\,
\alpha_{\; \nu}^+ \, \alpha_{\; \mu}) \ , \quad [p_\mu,\alpha^{\nu +}] =0\,.
\end{equation}
The first commutator in (\ref{COMU}) is nonzero if both $p_\mu$ and $p_\nu$
belong to the same Hilbert space.
 When acting on a vector $\xi_\mu$ the commutator (\ref{COMU})
has the form
\begin{equation} \label{Cnabla}
[ \nabla_\mu, \nabla_\nu] \xi_\rho = \frac{1}{L^2}(g_{\nu \rho}
\xi_\mu - g_{\mu \rho} \xi_\nu)\,,
\end{equation}
where $L$ is the radius of $AdS_{\cal D}$ space.
Again the expression (\ref{Cnabla}) is nonzero when $p_\mu, p_\nu$ and
$\xi_\rho$ are in the same Hilbert space.

In order to construct off--shell vertices one can
use the BRST charge for massless reducible representations
of the ${\cal D}$--dimensional AdS group \cite{Sagnotti:2003qa}
\begin{equation} \label{brstads}
Q =
c_0\hat{l}_0 +c l^{+}+c^+ l -\frac{8}{L^2} c_0(c^+b^+ M + bc M^+) - c^+cb_0,
\end{equation}
\begin{eqnarray}\label{l0AdS}
\hat{l}_0&=& {p} \cdot {p} +
\frac{1}{L^2}({(\alpha^{ +} \cdot \alpha)}^2 + {\cal
D}\alpha^{ +} \cdot \alpha -6 \alpha^{ +} \cdot \alpha -2{\cal
D} +6
-4 M^+ M +\nonumber \\
&& c^+b (4 \alpha^{ +} \cdot \alpha +2{\cal D} -6 ) + b^+c (4 \alpha^{ +} \cdot \alpha +2{\cal D} -6) +
12 c^+b b^+c).
\end{eqnarray}
For the off--shell description of irreducible massless higher spin modes one can use the corresponding BRST charge
\cite{Buchbinder:2001bs} (see also  \cite{Buchbinder:2007vq}--\cite{Grigoriev:2011gp}
 for BRST charges which correspond to various fermionic and bosonic representations of
  $AdS_{\cal D}$ group) which has more complicated form and includes more ghost variables than (\ref{brstads}).
In order to perform on-shell computations one can follow a similar line to that in the
previous Section for the case of Minkowski space--time.
At the same time
the massless higher spin fields and parameters of gauge transformations are restricted as \cite{Metsaev:1997nj}
\begin{eqnarray}
\label{AdStriplet2-1}
&& \Box \; \phi_{\mu_1,...,\mu_s} \ = \
  \frac{(2-s)(3-{\cal
D}-s)-s}{L^2} \,
\phi_{\mu_1,...,\mu_s}  \nonumber \ , \\
&&  \nabla^{\mu_1}  \phi_{\mu_1,...,\mu_s}=0   \ , \\
&&\phi^{\mu_1}{}_{\mu_1,...,\mu_s}=0, \nonumber
\end{eqnarray}
and
 \begin{eqnarray}
 \label{deltaphidads2-1}
&&\Box \ \Lambda_{\mu_1,...,\mu_{s-1}} \ = \ -\frac{(s-1)(3-s-{\cal D})}{L^2}\
   \Lambda_{\mu_1,...,\mu_{s-1}} \nonumber \, \\
  && \nabla^{\mu_1} \Lambda_{\mu_1,...,\mu_{s-1}} =0, \\
&&\Lambda^{\mu_1}{}_{\mu_1,...,\mu_{s-1}} =0. \nonumber
\end{eqnarray}
The conditions (\ref{AdStriplet2-1})--(\ref{deltaphidads2-1})
can be represented in an operatorial form similarly to the equations (\ref{EOM})--(\ref{0trace})
in  flat space
\begin{equation}\label{AdStriplet2-3}
({p} \cdot {p} +
\frac{1}{L^2}((\alpha^{ +} \cdot \alpha)^2 + {\cal
D}\alpha^{ +} \cdot \alpha -6 \alpha^{ +} \cdot \alpha -2{\cal
D} +6))^i |\phi_i \rangle = l^i|\phi_i \rangle= M^i|\phi_i \rangle=0,
\end{equation}
\begin{equation}\label{AdStriplet2-4}
({p} \cdot {p} +
\frac{1}{L^2}((\alpha^{ +} \cdot \alpha)^2 + {\cal
D}\alpha^{ +} \cdot \alpha -2 \alpha^{ +} \cdot \alpha))^i |\lambda_i \rangle = l^i|\lambda_i \rangle= M^i|\lambda_i \rangle=0,
\end{equation}
and the gauge transformation law is again
\begin{equation}\label{gaugeads}
\delta |\phi_i \rangle= l^{i+}|\lambda_i \rangle.
\end{equation}
As it was in the case for Minkowski space--time, in order to
construct cubic and quartic vertices on $AdS_{\cal D}$ one uses the equations
(\ref{quartons})--(\ref{brstinvons}). However, the fact that the operators $p_\mu$
do not commute makes the calculations more complicated.

In order to determine the cubic vertex on $AdS_{\cal D}$ one starts with the
solution on a Minkowski background (\ref{VERYTOTALTAYLOR}) and considers
the operators $p_\mu$ to be $AdS_{\cal D}$ covariant derivatives defined in
(\ref{pop}). In other words one makes an $AdS_{\cal D}$ deformation of
a flat vertex, similarly to how it has been done for the $s-0-0$
vertex in \cite{Fotopoulos:2007yq}.
For simplicity one can consider irreducible (traceless) representations
and ignore the $\Delta_2$ dependence in quartic and cubic vertices.

When considering the equation (\ref{brstinvons}) for the cubic vertex (the discussion for quartic vertices is
completely analogous) one can consider only expressions of the type
\begin{equation}
\langle \lambda_1|\langle \phi_2|\langle \phi_3| (\alpha^1\cdot p^1) \Delta_1^m \Delta_3^n |0\rangle_{123},
\end{equation}
and restore the rest of the equation using cyclic symmetry of the vertex.
As a first step
 one needs to
push the oscillators $\alpha^1_{\mu}$ from $l^1$ to the right until
they annihilate the vacuum.
The second step is to push the operators $p_\mu^1$
to the right of $ \Delta_1^m \Delta_3^n$ in order to be able to use the
total derivative condition $p_\mu^1 + p_\mu^2 +p_\mu^3=0$.
When performing these commutators the terms proportional
to $\frac{1}{L^2}$ appear.
In order to compensate them one adds  ``counterterms" to the vertex which are proportional
to $\frac{1}{L^2}$.
To this end first note that one can fix the total power in each term of the vertex
to be a constant $k$ in oscillators and solve the equation for this power \cite{Buchbinder:2006eq}.
The zeroth term in $\frac{1}{L^2}$ is a corresponding term in the flat vertex.
The first counterterm in $\frac{1}{L^2}$ is written by replacing the combination
$(\alpha^i\cdot p^j)^2\rightarrow \frac{1}{L^2}(\alpha^i \cdot \alpha^j)$ and so on.
Consider for example the term of the form
\begin{equation}
X_{i_1 j_1,i_2 j_2,i_3 j_3,i_4 j_4} (\alpha^{i_1} \cdot p^{j_1})(\alpha^{i_2} \cdot p^{j_2})(\alpha^{i_3} \cdot p^{j_3})(\alpha^{i_4} \cdot p^{j_4}),
\end{equation}
where $X_{i_1 j_1,i_2 j_2,i_3 j_3,i_4 j_4}$ is a set of coefficients. The first counterterm has a form
\begin{equation}
\frac{1}{L^2}Y_{i_1 j_1;m_1 n_1,m_2 n_2} (\alpha^{i_1} \cdot \alpha^{j_1})
(\alpha^{m_1} \cdot p^{n_1})(\alpha^{m_2} \cdot p^{n_2}),
\end{equation}
while the second counterterm has the form
\begin{equation}
\frac{1}{L^4}Z_{i_1 j_1,i_2 j_2} (\alpha^{i_1} \cdot \alpha^{j_1})(\alpha^{i_2} \cdot \alpha^{j_2})
\end{equation}
Finally, one uses the condition of the BRST invariance of the cubic vertex
 to transform the combinations $p^1\cdot(p^2-p^3)$ into
$(p^3)^2-(p^2)^2$, then pushes the latter combination to the left and  uses
the first of the constraints in (\ref{AdStriplet2-3})--(\ref{AdStriplet2-4}).

When moving momenta to the right one makes repeated use of the commutator

\begin{eqnarray}
(A \cdot p)^n p_\mu (A\cdot p) &=& (A\cdot p)^{n+1} p_\mu \nonumber \\
& &+\sum_{r=1}^{[\frac{n+1}{2}]}\frac{T_r}{L^{2r}}\binom{n}{2r-1}(A\cdot A)^{r-1}(A \cdot p)^{n-2r+1} \\
& &\hspace{30mm}\times\left[(A\cdot A)p_\mu -A_\mu (A\cdot p)\right], \nonumber
\end{eqnarray}

\noindent where $T_n$ are the \emph{tangent numbers} \cite{Knuth:1967}. Then, when moving momenta to the left, one uses the commutators

\begin{eqnarray}
(A \cdot p)^n p_\mu &=& p_\mu (A \cdot p)^n  \\ \nonumber
& & +\sum_{r=1}^{[\frac{n}{2}]}\frac{(-1)^{r-1}}{L^{2r}}\bn{n}{2r}(A \cdot A)^{r-1}
[(A \cdot p)\alpha_\mu-(A \cdot A)p_\mu ](A \cdot p)^{n-2r},
\end{eqnarray}
and
\begin{eqnarray}
(A \cdot p)^n p_\mu p^\mu &=& p_\mu p^\mu (A \cdot p)^n +\frac{{\cal D}-1}{L^2}\sum_{r=0}^{[\frac{n-1}{2}]}\frac{(-1)^r 2^{2r}}{L^{2r}}(A \cdot A)^r\binom{n}{2r+1}(A \cdot p)^{n-2r} \nonumber \\
& & +\sum_{r=1}^{[\frac{n}{2}]}\frac{(-1)^{r-1}\left(2^{2r-1}+1\right)}{L^{2r}}(A \cdot A)^{r-1}
\binom{n}{2r}  \nonumber \\
&&\hspace{10mm}\times p^\mu[(A \cdot p)A_\mu-(A \cdot A)p_\mu ](A \cdot p)^{n-2r}.
\end{eqnarray}

\noindent Here the $p_\mu$'s are assumed to be in the same Hilbert space, and $A$ represents any combination of oscillators and momenta in different Hilbert spaces.

\section{Conclusions}
In the present paper we considered a problem of constructing quartic vertices
for massless higher spin fields on a Minkowski background. Corresponding cubic vertices
are expressed in terms of one arbitrary function
as in (\ref{Delta1}) and (\ref{Delta2}) or in terms of two arbitrary functions as in
(\ref{1-3}).
We found that one can find a corresponding quartic vertex which restores
 gauge invariance of the Lagrangian (\ref{LIBRSTQ}) at the
$g^2$ order in coupling constant for any functions parametrizing the cubic vertex.

This fact, however, is not sufficient to conclude that the corresponding perturbation theory
is nontrivial already at the level of four point functions. In order to check if
four point scattering is nontrivial we considered  a ``four particle  test" suggested in \cite{Benincasa:2007xk}.
This test can be used for constructable theories, i.e. when the total four point function vanishes at complex infinity
under the BCFW shift.
We have taken the simplest case of a tree level amplitude where all four external particles are real scalars
interchanging higher spin modes as an internal propagator. This analysis shows that the constructable cubic vertices considered in this
paper do not pass the ``four particle test", which indicates the triviality of the corresponding S--matrix.
Similar results have been obtained in \cite{Fotopoulos:2010ay}
where a
 detailed analysis of four point amplitudes   leads to a suggestion
that the theory of massless higher spin fields on Minkowski background should be considered in a wider context, i.e.
one should add some nonlocal/composite objects to restore its consistency.

On the other side, the quartic vertices found in the present paper can be useful for studies
 of massless higher spin fields on a ${\cal D}$--dimensional anti-de Sitter
background. It is well known that the theory of massless higher spin fields on $AdS_{\cal D}$
is free from the troubles present in  analogous theories on Minkowski space.
One can therefore try to deform the cubic and quartic vertices on Minkowski space
to the ones on $AdS_{\cal D}$ (see \cite{Vasiliev:2011xf}  for an extensive discussion about construction
of cubic vertices on an $AdS_{\cal D}$ background from the cubic vertices on the Minkowski background)
using the technique presented in  Section 6. Let us note that, apart
 from the deformation of the vertices on the Minkowski space to $AdS_{\cal D}$ background,
 one also needs to be careful to take into account global symmetries of higher spin fields
 in order to make a precise connection with higher spin gauge theories on $AdS_{\cal D}$ \cite{Vasiliev:2011xf}.
The  consideration of quartic vertices on $AdS_{\cal D}$ seems to be interesting in its own right as well as
in the framework of the AdS/CFT correspondence (see for example \cite{Henneaux:2010xg} for recent progress in this direction).
\vspace{5mm}

\noindent {\bf Acknowledgments.} We are grateful to M. Vasiliev
for drawing our attention to consideration of quartic vertices for massless higher spin fields
and to T. Mohaupt for discussions.
The work of P.D. has been supported by a STFC grant
 ST/I505805/1. M.T. is grateful to his family who turned his stay in Auckland, New Zealand
 into a wonderful experience. \\

 \noindent {\bf Note added.} When the paper was in press we received a communication
from M.Taronna pointing out that the fact that one can find a quartic
vertex for arbitrary functions that parametrise the cubic vertices  in
(\ref{VERYTOTAL}) was established in \cite{Taronna:2011kt} using an alternative BRST analysis. We are
grateful to M.Taronna for drawing our attention to this result and for
an extensive discussion on the topics presented in our paper.

\renewcommand{\thesection}{A}

\setcounter{equation}{0}

\renewcommand{\theequation}{A.\arabic{equation}}

\section{Explicit calculations for $\Delta_1$ and $\Delta_2$ dependence}\label{Ap}
In this Appendix we present detailed calculations for the quartic vertex when the corresponding cubic vertex has the form
(\ref{Delta2}).
Again let us consider a term in the variation of the total Lagrangian, where the parameter of gauge transformations
is in the Hilbert space labeled by ``3". One has for the variation of the higher spin field
\begin{equation}
\delta_1|\phi_1\rangle = -a_1 g\langle\phi_2|\langle\lambda_3|F^{(1)}\left(Y_\alpha^+\right)e^{\Delta_2}|0\rangle_{123}.
\end{equation}
This variaton leads in turn  to the following variation of the cubic term in the Lagrangian
\begin{eqnarray}
\delta \tilde{L}
&=& a_1g^2\,{}_{123}\langle 0|F\left(Y_\alpha\right)\exp\left(a_2 M^1\right)
\exp\left(a_2\left(M^2 + M^3\right)\right)
|\phi_2\rangle|\phi_3\rangle  \\ \nonumber
& &\langle\phi_{2^\prime}|\langle\lambda_{3^\prime}|F^{(1)}\left(Y_\alpha^+\right)
\exp\left(a_2 M^{1+}\right)
\exp\left(a_2\left(M^{2^\prime +} + M^{3^\prime +}\right)\right)
|0\rangle_{12^\prime 3^\prime}.   \\ \nonumber
\label{eq:deltaL}
\end{eqnarray}
In order to  ``integrate out''  the Hilbert space  ``1'' one uses the following identities
\begin{enumerate}
\item When the operator  $M^{1+}$ acts on a  function $F(Y_\alpha)$ on the left one gets
\begin{eqnarray*}
{}_1\langle 0|F\left(Y_\alpha\right)\left(a_2 M^{1+}\right) &=&
{}_1\langle 0|F^{(2)}\left(Y_\alpha\right)\left[\frac{a_1^2 a_2}{2}p_{23}\cdot p_{23}\right] \\
&=&{}_1\langle 0|\left(-\frac{a_1^2 a_2}{2}s\right)F^{(2)}\left(Y_\alpha\right),
\end{eqnarray*}
\noindent and similarly when the  $M^1$ operator acts on a function $F(Y_{\alpha^+})$ on the right.
\item After one eliminates  all $M^1$ and $M^{1+}$ operators  in this way, one can use that, for any two functions $G(Y_\alpha)$ and $H(Y_\alpha^+)$
one has
\begin{equation*}
{}_1\langle 0|G\left(Y_\alpha\right)\alpha^{1+}\cdot\alpha^{1}
H\left(Y_\alpha^+\right)|0\rangle_1
= {}_1\langle0|G^{(1)}\left(Y_\alpha\right)H^{(1)}\left(Y_\alpha^+\right) \left(a_1^2 p_{23}\cdot p_{2^\prime 3^\prime}\right)|0\rangle_{1}.
\end{equation*}
\item For any two functions $P(Y_\alpha)$ and $Q(Y_\alpha^+)$ one has
\begin{align*}
{}_1\langle0|P\left(Y_\alpha\right)Q\left(Y_\alpha^+\right)|0\rangle_1
&=& {}_1\langle0|\sum_{k=0}^{\infty}\frac{1}{k!}\left(a_1^2 p_{23}\cdot p_{2^\prime 3^\prime}\right)^{k}P^{(k)}\!\left(2a_1\left(\alpha^2\cdot p_3-\alpha^3\cdot p_2\right)\right) \\
& &\quad\times Q^{(k)}\!\left(2a_1\left(\alpha^{2^\prime +}\cdot p_{3^\prime}-\alpha^{3^\prime +}\cdot p_{2^\prime}\right)\right)|0\rangle_1.
\end{align*}

\item
Finally one can explicitly evaluate the commutator
\begin{align}\nonumber
\left[\left(a_2 M_{1}\right)^{m},\left(a_2 M_{1}^{+}\right)^{n}\right] & = \sum_{k=1}^{\min(m,n)}\sum_{r=0}^{k}\left(a_2 M_{1}^{+}\right)^{n-k}\alpha_{\mu_{1}}^{1+}\ldots
\alpha_{\mu_{r}}^{1+}\binom{m}{k}\binom{n}{k}\binom{k}{r}k!a_2^{2k}\\ \nonumber
 &  \left(\frac{{\cal D}}{2}+m+n-\left(k+1\right)\right)
 \ldots\left(\frac{{\cal D}}{2}+m+n-\left(2k-r\right)\right) \\
 &\times \alpha_{\mu_{1}}^{1}\ldots\alpha_{\mu_{r}}^{1}\left(a_2 M\right)^{m-k}.
\end{align}

\end{enumerate}
Using these identities,  the part of \eqref{eq:deltaL} pertaining to the ``1'' Hilbert space becomes
\begin{eqnarray}
\delta\tilde{L} & \sim & {}_1\langle0|\sum_{m,n,s=0}^{\infty}\sum_{k=0}^{\min(m,n)}\sum_{r=0}^{k}\frac{k!}{m!n!s!}\binom{m}{k}\binom{n}{k}\binom{k}{r}a_2^{2k} \nonumber \\
 &  & \times\left(\frac{{\cal D}}{2}+m+n-\left(k+1\right)\right)\ldots\left(\frac{{\cal D}}{2}+m+n-\left(2k-r\right)\right) \nonumber \\
 &  & \times\left(-\frac{a_1^2 a_2}{2}s\right)^{n-k}\left(a_1^2 p_{23}\cdot p_{2^\prime 3^\prime}\right)^{r+s} \left(-\frac{a_1^2 a_2}{2}s\right)^{m-k} \nonumber \\
 & &\times F^{(2n-2k+s)}\!\left(2a_1\left(\alpha^2\cdot p_3-\alpha^3\cdot p_2\right)\right) \nonumber \\
  & &\times \exp\left(a_2\left(M^2+M^3+M^{2^\prime +}+M^{3^\prime +}\right)\right) \nonumber \\
  & &\times F^{(2m-2k+s+1)}\!\left(2a_1\left(\alpha^{2^\prime +}\cdot p_{3^\prime}-\alpha^{3^\prime +}\cdot p_{2^\prime}\right)\right)|0\rangle_1.
\end{eqnarray} In order to
simplify the above expression one can first
 keep $r$ constant and sum over all $k\geq r$, and $m,n\geq k$.
 Further,
 after these summations have been done
one can sum over all $r$. We first note that
\[
\frac{k!}{m!n!s!}\binom{m}{k}\binom{n}{k}\binom{k}{r} = \frac{1}{k!\left(m-k\right)!\left(n-k\right)!s!}\binom{k}{r}.
\]
Then, for example, for $r=0$, we get
\begin{eqnarray*}
\delta\tilde{L} & \sim & {}_1\langle 0|\sum_{m,n,s=0}^{\infty}\sum_{k=0}^{\infty}\frac{a_2^{2k}}{k!m!n!s!}\left(\frac{{\cal D}}{2}+m+n\right)\ldots\left(\frac{{\cal D}}{2}+m+n+k-1\right)\\
 &  & \times\left(-\frac{a_1^2 a_2}{2}s\right)^{m+n}\left(a_1^2 p_{23}\cdot p_{2^\prime 3^\prime}\right)^{s}
  F^{(2n+s)}\!\left(2a_1\left(\alpha^2\cdot p_3-\alpha^3\cdot p_2\right)\right) \\
  & &\times \exp\left(a_2\left(M^2+M^3+M^{2^\prime +}+M^{3^\prime +}\right)\right) \\
  & &\times F^{(2m+s+1)}\!\left(2a_1\left(\alpha^{2^\prime +}\cdot p_{3^\prime}-\alpha^{3^\prime +}\cdot p_{2^\prime}\right)\right)|0\rangle_1.
\end{eqnarray*}
One can now perform the sum over $k$, the only relevant parts of the
expression being
\[
\sum_{k=0}^{\infty}\frac{a_2^{2k}}{k!}X(X+1)\ldots\left(X+k-1\right),
\]
where we have put $X\equiv\frac{{\cal D}}{2}+m+n$. Now, the ``rising factorial''
in this expression can be written in terms of the \emph{Stirling numbers
of the first kind} as
\begin{equation}
X(X+1)\ldots(X+k-1)=\sum_{l=0}^{k}\left[{k\atop l}\right]X^{l}.
\end{equation}
Considering constant powers of $X^{l}$, we first sum over $k\geq l$, and then
over $l$ to obtain \cite{Abramowitz:1972}
\[
\sum_{l=0}^{\infty}X^{l}\sum_{k=l}^{\infty}\frac{a_2^{2k}}{k!}\left[{k\atop l}\right]=\exp\left[-X\log\left(1-a_2^{2}\right)\right]:=\exp\left(\beta X\right),
\]
\noindent where $\beta=-\log\left(1-a_2^{2}\right)$. Let us note that the singular case $a_2^2=1$
corresponds to the purely cubic theory \cite{Fotopoulos:2007nm}.
In fact, one can show that, for any fixed $r$, we have the sum
\[
\frac{a_2^{2r}}{r!}\sum_{k=0}^{\infty}\frac{a_2^{2k}}{k!}\left(X+r\right)\left(X+r+1\right)\ldots\left(X+r+k-1\right)=\frac{a_2^{2r}}{r!}\exp\left(\beta X+\beta r\right).
\]
Therefore, the whole expression simplifies to
\begin{eqnarray*}
\delta\tilde{L} & \sim & {}_1\langle 0|\sum_{m,n,r,s=0}^{\infty}\frac{1}{m!n!r!s!}a_2^{2r}\exp\left[\beta\left(\frac{{\cal D}}{2}+m+n+r\right)\right] \\
 &  & \left(-\frac{a_1^2 a_2}{2}s\right)^{m+n}\left(a_1^2 p_{23}\cdot p_{2^\prime 3^\prime}\right)^{r+s}
  F^{(2n+r+s)}\!\left(2a_1\left(\alpha^2\cdot p_3-\alpha^3\cdot p_2\right)\right) \\
  & &\times \exp\left(a_2\left(M^2+M^3+M^{2^\prime +}+M^{3^\prime +}\right)\right) \\
  & &\times F^{(2m+r+s+1)}\!\left(2a_1\left(\alpha^{2^\prime +}\cdot p_{3^\prime}-\alpha^{3^\prime +}\cdot p_{2^\prime}\right)\right)|0\rangle_1.
\end{eqnarray*}
Now, putting $s=t-r$, summing first over $r\leq t$ and then $t$, and using
\[
\sum_{r=0}^{t}\frac{1}{r!(t-r)!}\left(a_2^{2}e^{\beta}\right)^{r}=\frac{1}{t!}\left(1+a_2^{2}e^{\beta}\right)^{t}=\frac{1}{t!}\left(1-a_2^{2}\right)^{-t}=\frac{1}{t!}e^{\beta t},
\]
\noindent one finally gets
\begin{eqnarray}
\delta\tilde{L} & \sim & {}_1\langle 0|\sum_{m,n,t=0}^{\infty}\frac{1}{m!n!t!}  \left(-e^\beta\frac{a_1^2 a_2}{2}s\right)^{m+n}\left(e^\beta a_1^2 p_{23}\cdot p_{2^\prime 3^\prime}\right)^{t} \nonumber \\
  & &\times F^{(2n+t)}\!\left(2a_1\left(\alpha^2\cdot p_3-\alpha^3\cdot p_2\right)\right) \nonumber \\
  & &\times \exp\left(\frac{D}{2}\beta +a_2\left(M^2+M^3+M^{2^\prime +}+M^{3^\prime +}\right)\right) \\
  & &\times F^{(2m+t+1)}\!\left(2a_1\left(\alpha^{2^\prime +}\cdot p_{3^\prime}-\alpha^{3^\prime +}\cdot p_{2^\prime}\right)\right)|0\rangle_1. \nonumber
\end{eqnarray}

\end{document}